\documentclass[%
  twocolumn,
 showpacs,
 showkeys,
 preprintnumbers,
 amsmath,amssymb,
 aps,
  pra,
  longbibliography,
 ]{revtex4-1}

\usepackage[dvipsnames]{xcolor}

\usepackage{mathptmx}

\usepackage{amssymb,amsthm,amsmath}

\usepackage{tikz}
\usetikzlibrary{calc}

\usepackage[breaklinks=true,colorlinks=true,anchorcolor=blue,citecolor=blue,filecolor=blue,menucolor=blue,pagecolor=blue,urlcolor=blue,linkcolor=blue]{hyperref}
\usepackage{graphicx}
\usepackage{url}

\usepackage{xcolor}
\usepackage{colortbl}

\begin{document}

\title{Faithful orthogonal representations of graphs from partition logics}


\author{Karl Svozil}
\email{svozil@tuwien.ac.at}
\homepage{http://tph.tuwien.ac.at/~svozil}

\affiliation{Institute for Theoretical Physics,
Vienna  University of Technology,
Wiedner Hauptstrasse 8-10/136,
1040 Vienna,  Austria}

\date{\today}

\begin{abstract}
The graphs induced by partition logics allow a dual probabilistic interpretation: a classical one for which probabilities lie on the convex hull of the dispersion-free weights, and another one, suggested independently from the quantum Born rule, in which probabilities are formed by the (absolute) square of the inner product of state vectors with the faithful orthogonal representations of the respective graph. Two immediate consequences are the demonstration that the logico-empirical structure of observables does not determine the type of probabilities alone, and that complementarity does not imply contextuality.
\end{abstract}


\maketitle

\section{Partition logics as nonboolean structures pasted from Boolean subalgebras}

Partitions provide ways to distinguish between elements of a given finite set ${\cal S}_n=\{1,2,\ldots ,n\}$.
The Bell number $B_n$ (after Eric Temple Bell) is the number of such partitions~\cite{Sloane_oeis.org/A000110}.
(Obvious generalizations are infinite denumerable sets or continua.)
We shall restrict our attention to partitions with an equal number $1\le m\le n$ of elements.
Every partition can be identified with some Boolean subalgebra $2^m$ -- in graph theoretical terms a clique --
of $2^n$ whose atoms are the elements of that partition.

A partition logic~\cite{svozil-93,schaller-92,dvur-pul-svo,schaller-96,svozil-2001-eua} is  the logic obtained
(i) from collections of such partitions, each partition being identified with an $m$-atomic Boolean subalgebra of $2^n$, and
(ii) by ``stitching'' or pasting these subalgebras through identifying identical intertwining elements.
In quantum logic this is referred to as pasting construction; and the partitions are identified with, or
are synonymously denoted by, blocks, subalgebras or cliques, which are representable by orthonormal bases or maximal operators.

Partitions represent classical mini-universes which satisfy compatible orthogonality, or
Specker's exclusivity principle~\cite{specker-60,specker-ep,Cabello-2013-SimpleExpl,fritz-2013,Henson-2012,Cabello-2012-SpeckersPrinciple,cabello-2013-beg,Cabello-2014-gtatqc}:
if any two observables corresponding to two elements of a partition are co-measurable, the entire set of observables corresponding to all
elements of that partition are simultaneously measurable.
(For Hilbert spaces this is a well-known theorem; see, for instance, von Neumann~\cite[Satz~8, p.~221]{v-neumann-31}
and~\cite[p.~173]{v-neumann-55}, or Halmos~\cite[\S~84, Theorem~1, p.~171]{halmos-vs}.)

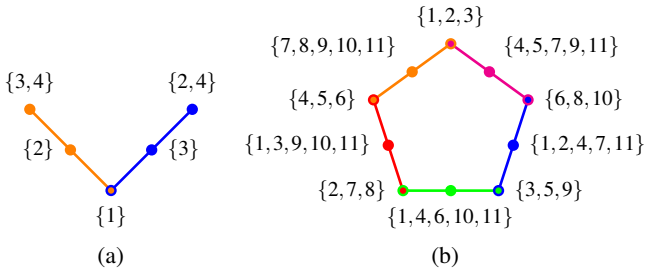
\begin{figure}
\begin{center}
\begin{tabular}{cc}
\begin{tikzpicture}  [scale=0.18]

\newdimen\ms
\ms=0.05cm

\tikzstyle{every path}=[line width=1pt]

\tikzstyle{c3}=[circle,inner sep={\ms/8},minimum size=3*\ms]
\tikzstyle{c2}=[circle,inner sep={\ms/8},minimum size=1.7*\ms]
\tikzstyle{c1}=[circle,inner sep={\ms/8},minimum size=0.8*\ms]

\newdimen\R
\R=6cm     



\path
  (0,6 ) coordinate(1)
  (3,3    ) coordinate(2)
  (6,0 ) coordinate(3)
  (9,3) coordinate(4)
  (12,6  ) coordinate(5)
;


\draw [color=orange] (1) -- (2) -- (3);
\draw [color=blue] (3) -- (4) -- (5);

%
%
\draw (1) coordinate[c3,fill=orange,label={above:\footnotesize $\{3,4\}$}];   %
\draw (2) coordinate[c3,fill=orange,label={left:\footnotesize $\{2\}$}];    %
\draw (3) coordinate[c3,fill=blue,label={below:\footnotesize $\{1\}$}]; %
\draw (3) coordinate[c2,fill=orange];  %
\draw (4) coordinate[c3,fill=blue,label={right:\footnotesize $\{3\}$}];  %
\draw (5) coordinate[c3,fill=blue,label={above:\footnotesize $\{2,4\}$}];  %

\end{tikzpicture}
&
\begin{tikzpicture}  [scale=0.18]

\newdimen\ms
\ms=0.05cm

\tikzstyle{every path}=[line width=1pt]

\tikzstyle{c3}=[circle,inner sep={\ms/8},minimum size=3*\ms]
\tikzstyle{c2}=[circle,inner sep={\ms/8},minimum size=1.7*\ms]
\tikzstyle{c1}=[circle,inner sep={\ms/8},minimum size=0.8*\ms]

\newdimen\R
\R=6cm     



\path
  ({90 + 0 * 360 /5}:\R      ) coordinate(1)
  ({90 + 36 + 0 * 360 /5}:{\R * sqrt((25+10*sqrt(5))/(50+10*sqrt(5)))}      ) coordinate(2)
  ({90 + 1 * 360 /5}:\R   ) coordinate(3)
  ({90 + 36 + 1 * 360 /5}:{\R * sqrt((25+10*sqrt(5))/(50+10*sqrt(5)))}   ) coordinate(4)
  ({90 + 2 * 360 /5}:\R  ) coordinate(5)
  ({90 + 36 + 2 * 360 /5}:{\R * sqrt((25+10*sqrt(5))/(50+10*sqrt(5)))}  ) coordinate(6)
  ({90 + 3 * 360 /5}:\R  ) coordinate(7)
  ({90 + 36 + 3 * 360 /5}:{\R * sqrt((25+10*sqrt(5))/(50+10*sqrt(5)))}  ) coordinate(8)
  ({90 + 4 * 360 /5}:\R     ) coordinate(9)
  ({90 + 36 + 4 * 360 /5}:{\R * sqrt((25+10*sqrt(5))/(50+10*sqrt(5)))}     ) coordinate(10)
;


\draw [color=orange] (1) -- (2) -- (3);
\draw [color=red] (3) -- (4) -- (5);
\draw [color=green] (5) -- (6) -- (7);
\draw [color=blue] (7) -- (8) -- (9);
\draw [color=magenta] (9) -- (10) -- (1);    %

%
%
\draw (1) coordinate[c3,fill=orange,label=90:{\footnotesize $\{ 1,2,3\} $}];   %
\draw (1) coordinate[c2,fill=magenta];  %
\draw (2) coordinate[c3,fill=orange,label={above left:\footnotesize $\{ 7,8,9,10,11\}$}];    %
\draw (3) coordinate[c3,fill=red,label={left:\footnotesize $\{ 4,5,6\} $}]; %
\draw (3) coordinate[c2,fill=orange];  %
\draw (4) coordinate[c3,fill=red,label={left:\footnotesize $\{ 1,3,9,10,11\}$}];  %
\draw (5) coordinate[c3,fill=green,label={left:\footnotesize $\{ 2,7,8\} $}];  %
\draw (5) coordinate[c2,fill=red];  %
\draw (6) coordinate[c3,fill=green,label={below:\footnotesize $\{ 1,4,6,10,11\} $}];
\draw (7) coordinate[c3,fill=blue,label={right:\footnotesize $\{ 3,5,9\}$}];  %
\draw (7) coordinate[c2,fill=green];  %
\draw (8) coordinate[c3,fill=blue,label={right:\footnotesize $\{ 1,2,4,7,11\}$}];  %
\draw (9) coordinate[c3,fill=magenta,label={right:\footnotesize $\{ 6,8,10\}$}];
\draw (9) coordinate[c2,fill=blue];  %
\draw (10) coordinate[c3,fill=magenta,label={above right:\footnotesize $\{ 4,5,7,9,11\}$}];  %
\end{tikzpicture}
\\
(a)&(b)
\end{tabular}
\end{center}
\caption{Greechie orthogonality diagrams of (a) the $L_{12}$ logic and (b) the pentagon (pentagram) logic,
with two of their associated (quasi)classical partition logic representations.
}
\label{2018-q-f1}
\end{figure}

Unlike complete graphs $K_m$ representations of $m$-vertex cliques in which every pair of distinct vertices is connected by a unique edge
in quantum logic it is quite common to conveniently depict cliques (aka contexts) as smooth curves;
referred to as Greechie orthogonality diagram~\cite{greechie:71}.
For example, the two {\it ad hoc} partitions
\begin{equation}
\{\{1\},\{2\},\{3,4\}\} \text{ and }
\{\{1\},\{3\},\{2,4\}\}
\label{2018-e-pll12}
\end{equation}
of ${\cal S}_3=\{1,2,3,4\}$ form two 3-atomic Boolean algebras $2^3$ with  one identical intertwining atom $\{1\}$, as depicted in depicted in Fig.~\ref{2018-q-f1}(a).
It is the logic $L_{12}$ (because it has 12 elements).
-- is just two straight lines (representing the two contexts or cliques) interconnected at $\{1\}$.
(Cf. Fig.~1 of Wright's ``Bowtie Example 3.1''~\cite[pp.~884,885]{wright}.)

Many partition logics, such as the pentagon logic, have quantum doubles.
One of the (necessary and sufficient) criteria for quantum logics
to be representable by a partition logic is the separability of pairs of atoms of the logic by dispersion free
(aka two-valued, $\{0,1\}$-valued) weights/states~\cite[Theorem~0, p.~67]{kochen1}, interpretable as classical truth assignments.

Conversely, ``sufficiently'' many (more precisely, a separating set of) dispersion free states allows the
explicitly (re)construction of a partition logic~\cite{svozil-2001-eua,svozil-2008-ql}.
For instance, the five cyclically intertwined contexts/cliques forming a pentagon/pentagram logic~\cite[p.~267, Fig.~2]{wright:pent}
support 11 dispersion free states $v_1, \ldots , v_{11}$. By constructing 5 contexts/cliques from the occurrences of the dispersion free value $1$ on the respective 10 atoms
results in the partition logic based on the set of indices of the dispersion free states ${\cal S}_{11}=\{1,\ldots ,11\}$, as depicted in Fig.~\ref{2018-q-f1}(b)
\begin{equation}
\begin{aligned}
&\{
 \{ \{ 1,2,3\} ,\{ 4,5,7,9,11\} ,\{ 6,8,10\} \} ,\\
&\{ \{ 6,8,10\} ,\{ 1,2,4,7,11\} ,\{ 3,5,9\} \} ,\\
&\{ \{ 3,5,9\} ,\{ 1,4,6,10,11\} ,\{ 2,7,8\} \} ,\\
&\{ \{ 2,7,8\} ,\{ 1,3,9,10,11\} ,\{ 4,5,6\} \} ,\\
&\{ \{ 4,5,6\} ,\{ 7,8,9,10,11\} ,\{ 1,2,3\} \}
\}.
\end{aligned}
\label{2018-e-plpentagon}
\end{equation}
According to this construction, the earlier logic $L_{12}$  would have a partition logic representation
$\{
\{ \{2,3\},\{4,5\},\{1\} \},
\{ \{1\},\{3,5\},\{2,4\} \}
\}$
based on ${\cal S}_{5}=\{1,\ldots ,5\}$, corresponding to its 5 dispersion free states.

\section{Probabilities on partition logics}

The following hypothesis or principle is taken for granted:
{\em probabilities and expectations on classical substructures of an empirical logic
should be classical.}
That is,
mutually exclusive co-measurable propositions (satisfying Specker's exclusivity principle)
should obey the Kolmogorovian axioms; in particular, nonnegativity and additivity.
Nonnegativity implies that all probabilities are nonnegative: $P(E_1),\ldots,P(E_k)\ge 0$.
Additivity among (pairwise) mutually exclusive outcomes $E_1,\ldots , E_k$ means that
the probabilities of joint outcomes are equal to the sum of probabilities of these outcomes; that is,
within cliques/contexts, for $k\le m$:
$P(E_1\vee \cdots \vee E_k) =P(E_1)+\cdots + P(E_k) \le 1$.
In particular, probabilities add to 1 on each of the cliques/contexts.

At the moment at least three such types of probabilities are known to satisfy Specker's exclusivity principle, corresponding to
classical, quantum and  Wright's ``exotic''pure weights, such as the weight $\frac{1}{2}$ on the vertices of the pentagon~\cite[$\omega_0$, p.~68]{wright:pent}
and on the triangle vertices~\cite[pp.~899-902]{wright} (the latter logic is representable as partition logic~\cite[Example~8.2, pp.~420,421]{dvur-pul-svo}, but not in 3-dimensional Hilbert space).
The former two ``nonexotic'' types, based on representations of mutually disjoint sets, and on mutually orthogonal vectors, will be discussed later.

It is not too difficult to imagine boxes allowing input/output analysis ``containing'' classical or quantum algorithms, agents or mechanisms rendering the desired properties.
For instance, a model realization of a classical box rendering classical probabilities is Wright's generalized urn model~\cite{wright:pent,wright,svozil-2005-ln1e,2010-qchocolate}
or the initial state identification problem for  finite deterministic automaton~\cite{e-f-moore,svozil-93,schaller-95,schaller-96} --
both are equivalent models of partition logics~\cite{svozil-2001-eua} featuring complementarity without value indefiniteness.

{S}pecker's parable of the overprotective seer~\cite{specker-60,LIANG20111,LIANG2017110,svozil-2016-s} involving three boxes
is an example for which the exclusivity principle does {\em not} hold~\cite[Section~116, p.~40]{Tarrida-2014}.
It is an interesting problem to find other potential probability measures based on different approaches which are also linear in mutually exclusive events.

\subsection{Probabilities from the convex hull of dispersion-free states}
\label{2018-g-polyhedral}

For nonboolean logics, it is not immediately evident which probability measures should be chosen.
The answer is already implicit in  Zierler  and Schlessinger's  1965 paper
on {\em ``Boolean embeddings of orthomodular sets and quantum logic''}.
Theorem~0 of Kochen and Specker's 1967 paper~\cite{kochen1} states that separability by dispersion free states (of image $2^1={0,1}$)
for every pair of atoms of the lattice is a necessary and sufficient criterium for a homomorphic embedding into some ``larger'' Boolean algebra.
In 1978 Wright
explicitly stated~\cite[p.~272]{wright:pent} {\em ``that every urn weight is ``classical'', i.e., in the convex hull of the dispersion free weights''.}
In the graph theoretical context Gr{\"o}tschel, Lov{\'a}sz and  Schrijver
have discussed the {\em vertex packing polytope} $VP(G)$ of a graph $G$, defined as the convex hull
of incidence vectors of independent sets of nodes~\cite{GroetschelLovaszSchrijver1986}.
This author has employed dispersion free weights for hull computations on the Specker bug~\cite{svozil-2001-cesena} and other (partition) logics supporting
a separating set of two-value states.

Hull computations based on the pentagon (modulo pentagon/pentagram graph isomorphisms) can be found in Refs.~\cite{Klyachko-2008,Bub-2009,Bub-2010,Badziag-2011}
(for a survey see \cite[Section~12.9.8.3]{svozil-pac}).
The Bub and Stairs inequality~\cite[Equation~(10), p.~697]{Bub-2009} can be directly read off from the partition logic ~(\ref{2018-e-plpentagon}), as depicted in Fig.~\ref{2018-q-f1}(b),
which in turm are the cumulated indices of the nonzero dispersion free weights on the atoms:
the sum of the convex hull of the dispersion free weights on the 5 intertwining atoms (the ``vertices'' of the pentagon diagram) represented by the subsets
$\{1,2,3\}    $,
$\{ 6,8,10\}$,
$\{ 3,5,9\} $,
$\{ 2,7,8\} $,
$\{ 4,5,6\} $
of ${\cal S}_{11}$
are
\begin{equation}
\begin{aligned}
&
(\lambda_1+\lambda_2+\lambda_3   )+
(\lambda_6+\lambda_8+\lambda_{10}  )+
(\lambda_3+\lambda_5+\lambda_9   )+ \\
&\quad +
(\lambda_2+\lambda_7+\lambda_8   )+
(\lambda_4+\lambda_5+\lambda_6   )
\le 2 \sum_{i=1}^{11} \lambda_i =2
.
\end{aligned}
\label{2018-e-bs}
\end{equation}

\subsection{Born-Gleason-Gr\"otschel-Lov\'asz-Schrijver type probabilities}

Motivated by cryptographic issues outside quantum theory Lov\'asz~\cite{lovasz-79} has proposed an ``indexing'' of vertices of a graph
by vectors reflecting their adjacency: the graph-theoretic definition of a faithful orthogonal representation of a graph is by identifying vertices with vectors
(of some Hilbert space of dimension $d$) such that any pair of vectors are orthogonal if and only if their vertices are {\em not} orthogonal~\cite{lovasz-89,Parsons-1989}.
For physical applications, Cabello\cite{Cabello-2010-ncoptaa}, Portillo~\cite{Portillo-2015} and others have
uses an ``inverse'' notation, in which vectors are required to be mutually orthogonal whenever they are adjacent.
Both notations are equivalent by exchanging graphs with their complements or inverses.

There is no systematic algorithm to compute the minimal dimension for a faithful orthogonal representation of a graph.
Lov\'asz~\cite{lovasz-79,cabello-2013-beg} gave a (relative to entropy measures~\cite{Haemers-1979}) ``optimal'' vector representation of the pentagon
graph depicted in Figure~\ref{2018-q-f1}(b) in three dimensions [$L_{12}$ depicted in Figure~\ref{2018-q-f1}(a) is a sublogic thereof]:
modulo pentagon/pentagram graph isomorphisms which, in two-line notation is $\begin{pmatrix}
1&2&3&4&5\\
1&4&2&5&3
\end{pmatrix}$, and in cycle notation is $(1)(2453)$,
its set of five intertwining vertices $\{v_1,\ldots ,v_5\}=\{u_1,u_3,u_5,u_2,u_4\}$  are
represented by the 3-dimensional unit vectors (the five vectors corresponding to the ``inner'' vertices/atoms can be found by a Gram-Schmidt process)
\begin{equation}
\vert u_l \rangle = 5^{-\frac{1}{4}}
\begin{pmatrix}
1,
\left[5^\frac{1}{2}-1\right]^\frac{1}{2}
\cos  \frac{2 \pi l}{5}
,
\left[5^\frac{1}{2}-1\right]^\frac{1}{2}
\sin  \frac{2 \pi l}{5}
\end{pmatrix}
,
\end{equation}
which, by preparing the ``(umbrella) handle'' state vector $\begin{pmatrix}
1,0,0
\end{pmatrix} $, turns out to render the maximal~\cite{Bub-2009,Badziag-2011} quantum bound $\sum_{j=1}^5 \langle c\vert u_l \rangle^2
=\sqrt{5}$,
which exceeds the ``classical'' bound~(\ref{2018-e-bs}) of $2$ from the computation of the
convex hull of the dispersion free weights.

Based on Lov\'asz's vector representation by graphs
Gr{\"o}tschel, Lov{\'a}sz and  Schrijver have proposed~\cite[Section~3]{GroetschelLovaszSchrijver1986}
a Gleason-Born type probability measure~\cite{Cabello-2018-BornRule}
which results in convex sets different from polyhedra defined {\it via} convex hulls of vectors discussed earlier in Section~\ref{2018-g-polyhedral}.
Essentially their probability measure is based upon
$m$-dimensional faithful orthogonal representations of a graph $G$
whose vertices $v_i$ are represented by unit vectors $\vert v_i\rangle$  which are orthogonal within,
and nonorthogonal outside, of cliques/contexts.
Every  vertex $v_i$ of the graph $G$, represented by the unit vector $\vert v_i\rangle$,
can then be associated with a ``probability'' with respect to
some unit ``preparation'' (state) vector $\vert c\rangle$
by defining this ``probability'' to be the square of the inner product of  $\vert v_i\rangle$ and  $\vert c\rangle$;
that is,  by
$P(c,v_i)=\langle c \vert v_i\rangle^2$.
Iff the vector representation (in the sense of Cabello-Portillo) of $G$ is faithful
the Pythagorean theorem assures that, within every clique/context of $G$,
probabilities are positive, additive, and  (as both $\vert v_i\rangle$ and  $\vert c\rangle$ are normalized)
the sum of probabilities
on that context
adds up to exactly one; that is,
$\sum_{i \in \text{clique/context}} P(c,v_i)=1$.
Thereby, probabilities and expectations
of simultaneously comeasurable observables, represented by graph vertices within cliques or contexts,
obey Specker's exclusivity principle and ``behave classically''.
It might be challenging to motivate ``quantum type'' probabilities and their convex expansion, the theta body~\cite{GroetschelLovaszSchrijver1986},
by the very assumptions such as exclusivity~\cite{Cabello-2014-gtatqc,Cabello-2018-BornRule}.

A very similar measure on the closed subspaces of Hilbert space,
satisfying Specker's exclusivity principle and additivity,
had been proposed by Gleason~\cite[first and second paragraphs, p.~885]{Gleason}:
{\em ``A measure on the closed
subspaces means a function $\mu$ which assigns to every closed subspace a nonnegative
real number such that if $\{A_i\}$ is a countable collection of mutually
orthogonal subspaces having closed linear span $B$, then
$\mu (B) = \sum_i \mu(A_i)$.
It is easy to see that such a measure can be obtained by selecting a vector $v$
and, for each closed subspace $A$, taking $\mu (A)$ as the square of the norm of the
projection of $v$ on $A$.''}
Gleason's derivation of the quantum mechanical Born rule~\cite[Footnote~1, Anmerkung bei der Korrektur, p.~865]{born-26-1}
operates in dimensions higher than two, and allows also mixed states;
that is, outcomes of nonideal measurements.
However, mixed states can always be ``completed''
or ``purified''~\cite[Section~2.5,pp.~109-111]{nielsen-book10}
(and thus outcomes of nonideal measurements made ideal~\cite{Cabello-2018-BornRule}) by
the inclusion of auxiliary dimensions.

\section{Quasiclassical analogues of entanglement}

In what follows classical analogs to entangled states will be discussed.
These examples are local.
They are based on Schr\"odinger's observation that entanglement among pairs of particles is associated with, or at least accompanied by,
{\em joint} or {\em relational}~\cite{zeil-99} properties of the constituents,
whereas nonentangled states feature individual separate properties of the pair constituents~\cite{schrodinger,CambridgeJournals:1737068,CambridgeJournals:2027212}.
(For early similar discussions in the measurement context,
see von Neumann~\cite[Section~VI.2, p~426, pp~436-437]{v-neumann-55} and London and Bauer~\cite{london-Bauer-1939,london-Bauer-1983}.)

\subsection{Partitioning of state space}
\label{2018-g-poss}

Wrights generalized urn model~\cite{wright:pent,wright}, in a nutshell, is the observation of black balls, on which
multiple colored symbols are painted, with monochromatic filters in only one of those colors.
Complementarity manifests itself in the necessity of choice of the particular color one observes: one may thereby obtain knowledge of the information encoded in this color;
but thereby invariable loses messages encoded in different colors.
A typical example is the logic $L_{12}$  encoded by the partition logic enumerated in~(\ref{2018-e-pll12}) and depicted in Figure~\ref{2018-q-f1}(a):
suppose that there are 4 ball types and two colors on black backgrounds:
\begin{itemize}
\item
ball type 1 is colored with orange a and blue a;
\item
ball type 2 is colored with orange b and blue c;
\item
ball type 3 is colored with orange c and blue b;
\item
ball type 4 is colored with orange c and blue c.
\end{itemize}

Suppose an urn is loaded with balls of all four types.
Suppose further that an agent's task is
to draw one ball from the urn and, by observing this ball, to find which type it is.
Of course, if the observer is allowed to look at both colors simultaneously, this would allow to single out exactly one ball type.
But that maximal resolution can no longer be maintained
in experiments restricted to one of the two colors.
Any one of such two experiment varieties could resolve different, complementary properties:
looking at the drawn ball with orange glasses the agent is able to resolve between balls
(i) of type 1 associated with the symbol a;
(ii) of type 2 associated with the symbol b; and
(ii) of type 3 or 4 associated with the symbol c.
The resolution between type 3 and 4 balls is lost.
Alternatively, by
looking at the drawn ball with blue glasses the agent is able to resolve between balls
(i) of type 1 associated with the symbol a;
(ii) of type 3 associated with the symbol b; and
(ii) of type 2 or 4 associated with the symbol c.
That is, for the color blue the resolution between type 2 and 4 balls is lost.
In any case, the state of ball type is partitioned in different ways, depending on the color of the filter.
(Similar considerations apply for initial state identification problems on finite automata.)

\subsection{Relational encoding}

Tables~\ref{2018-g-tp1} and~\ref{2018-g-tp2} enumerate a relational encodings among two or more colors
not dissimilar to Peres' detonating bomb model~\cite{peres222}.
Suppose that an urn is loaded with balls of the type occurring in subensemble $E_6$
of Table~\ref{2018-g-tp1}.
The observation of some symbol $s\in \{0,1\}$ in green implies the (counterfactual) observation of the same symbol $s$ in red,
and {\it vice versa}.
Table~\ref{2018-g-tp2} is just an extension to two colors per observer,
and an urn loaded with subensembles $(E_6)^2$
Agent Alice draws a ball from the urn and looks at it with her red (exclusive) or blue filters. Then Alice hands the ball over to Bob.
Agent Bob looks at the ball with his green (exclusive) or orange filters.
This latter scenario is similar to a Clauser-Horne-Shimony-Holt scenario of the Einstein-Podolsky-Rosen type,
except that the former is totally local and its probabilities derived from the convex hull of the dispersion-free weights
can never violate classical bounds; whereas
the latter one may be (and hopefully is) nonlocal,
and its performance with a quantum resource violates the classical bounds.

\begin{table}
\begin{ruledtabular}
\begin{tabular}{ccccccc}

sample & ball type 1  & ball type 2  & ball type 3  & ball type 4    \\

\hline

$E_1$ &  {\color{red}0}{\color{green}0} &  {\color{red}0}{\color{green}1} &   &   \\
$E_2$ &   &  &  {\color{red}1}{\color{green}0} &  {\color{red}1}{\color{green}1} \\
$E_3$ &  {\color{red}0}{\color{green}0} &   &  {\color{red}1}{\color{green}0} &   \\
$E_4$ &   &  {\color{red}0}{\color{green}1} &    &  {\color{red}1}{\color{green}1} \\

\hline

$E_5$ & {\color{red}0}{\color{green}0} &   &   &  {\color{red}1}{\color{green}1} \\
$E_6$ &   &  {\color{red}0}{\color{green}1} &  {\color{red}1}{\color{green}0} &

\end{tabular}
\end{ruledtabular}

\caption{Six subensembles $E_1$--$E_6$ of the set
$\{
 {\color{red}0}{\color{green}0} ,
 {\color{red}0}{\color{green}1} ,
 {\color{red}1}{\color{green}0} ,
 {\color{red}1}{\color{green}1}
\}$ with the following properties:
$E_1 = \{
 {\color{red}0}{\color{green}0} ,
 {\color{red}0}{\color{green}1}
\}$
encodes the first digit being $0$;
$E_2 = \{
 {\color{red}1}{\color{green}0} ,
 {\color{red}1}{\color{green}1}
\}$
encodes the first digit being $1$;
$E_3 = \{
 {\color{red}0}{\color{green}0} ,
 {\color{red}1}{\color{green}0}
\}$
encodes the second digit being $0$;
$E_4 = \{
 {\color{red}0}{\color{green}1} ,
 {\color{red}1}{\color{green}1}
\}$
encodes the second digit being $1$;
$E_5 = \{
 {\color{red}0}{\color{green}0} ,
 {\color{red}1}{\color{green}1}
\}$
encodes the first and the second digit being equal;
$E_6 = \{
 {\color{red}0}{\color{green}1} ,
 {\color{red}1}{\color{green}0}
\}$
encodes the first  and the second digit being different.
\label{2018-g-tp1} }
\end{table}

\begin{table}
\begin{ruledtabular}
\begin{tabular}{ccccccc}

sample & ball type 1  & ball type 2  & ball type 3  & ball type 4    \\

\hline

$(E_5)^2$ &  {\color{red}0}{\color{green}0}{\color{blue}0}{\color{orange}0} &
 {\color{red}0}{\color{green}0}{\color{blue}1}{\color{orange}1} &
 {\color{red}1}{\color{green}1}{\color{blue}0}{\color{orange}0} &
 {\color{red}1}{\color{green}1}{\color{blue}1}{\color{orange}1} \\
$(E_6)^2$ &  {\color{red}0}{\color{green}1}{\color{blue}0}{\color{orange}1}&
 {\color{red}0}{\color{green}1}{\color{blue}1}{\color{orange}0} &
 {\color{red}1}{\color{green}0}{\color{blue}0}{\color{orange}1} &
 {\color{red}1}{\color{green}0}{\color{blue}1}{\color{orange}0}

\end{tabular}
\end{ruledtabular}

\caption{The subensembles $(E_5)^2$ and $(E_6)^2$ of the set
$\{
 {\color{red}0}{\color{green}0}{\color{blue}0}{\color{orange}0}$,
${\color{red}0}{\color{green}0}{\color{blue}0}{\color{orange}1}$,
${\color{red}0}{\color{green}0}{\color{blue}1}{\color{orange}0}$,
${\color{red}0}{\color{green}0}{\color{blue}1}{\color{orange}1}$,
${\color{red}0}{\color{green}1}{\color{blue}0}{\color{orange}0}$,
${\color{red}0}{\color{green}1}{\color{blue}0}{\color{orange}1}$,
${\color{red}0}{\color{green}1}{\color{blue}1}{\color{orange}0}$,
${\color{red}0}{\color{green}1}{\color{blue}1}{\color{orange}1}$,
${\color{red}1}{\color{green}0}{\color{blue}0}{\color{orange}0}$,
${\color{red}1}{\color{green}0}{\color{blue}0}{\color{orange}1}$,
${\color{red}1}{\color{green}0}{\color{blue}1}{\color{orange}0}$,
${\color{red}1}{\color{green}0}{\color{blue}1}{\color{orange}1}$,
${\color{red}1}{\color{green}1}{\color{blue}0}{\color{orange}0}$,
${\color{red}1}{\color{green}1}{\color{blue}0}{\color{orange}1}$,
${\color{red}1}{\color{green}1}{\color{blue}1}{\color{orange}0}$,
${\color{red}1}{\color{green}1}{\color{blue}1}{\color{orange}1}
\}$ with the following properties:
$E_5 = \{
 {\color{red}0}{\color{green}0}{\color{blue}0}{\color{orange}0},
 {\color{red}0}{\color{green}0}{\color{blue}1}{\color{orange}1},
 {\color{red}1}{\color{green}1}{\color{blue}0}{\color{orange}0},
 {\color{red}1}{\color{green}1}{\color{blue}1}{\color{orange}1}
\}$
encodes the first and the second pair, as well as the third and the fourth pair of digits being equal;
$E_6 = \{
 {\color{red}0}{\color{green}1}{\color{blue}0}{\color{orange}1},
 {\color{red}0}{\color{green}1}{\color{blue}1}{\color{orange}0},
 {\color{red}1}{\color{green}0}{\color{blue}0}{\color{orange}1},
 {\color{red}1}{\color{green}0}{\color{blue}1}{\color{orange}0}
\}$
encodes the first  and the second pair, as well as the third and the fourth pair of digits being different.
\label{2018-g-tp2} }
\end{table}

\section{Partition logic freak show}

\begin{figure}
\begin{center}
\begin{tabular}{cc}
\begin{tikzpicture}  [scale=0.18]

\newdimen\ms
\ms=0.05cm

\tikzstyle{c3}=[circle,inner sep={\ms/8},minimum size=3*\ms]
\tikzstyle{c2}=[circle,inner sep={\ms/8},minimum size=1.7*\ms]
\tikzstyle{c1}=[circle,inner sep={\ms/8},minimum size=0.8*\ms]

\tikzstyle{every path}=[line width=1pt]

\newdimen\R
\R=6cm     



\path
  ({90 + 0 * 360 /3}:\R      ) coordinate(1)
  ({90 + 1 * 360 /3}:\R   ) coordinate(3)
  ({90 + 2 * 360 /3}:\R  ) coordinate(5)
;



\draw [color=orange] (1) -- (3);
\draw [color=blue] (3) -- (5);
\draw [color=green] (5) -- (1);

%
%

\draw (1) coordinate[c3,fill=orange];   %
\draw (1) coordinate[c2,fill=green,label=90:\colorbox{white}{\footnotesize  $\{1\}$}];  %
\node[c3,fill=orange,label={left:\colorbox{white}{\footnotesize   $\{2,4\}$}}] at ( $ (1)!{1/2}!(3) $ ) (2) {};    %
\draw (3) coordinate[c3,fill=orange];  %
\draw (3) coordinate[c2,fill=blue,label=below left:{\footnotesize  \colorbox{white}{$\{3\}$}}];  %
\node[c3,fill=blue,label={below:{\footnotesize  \colorbox{white}{$\{1,4\}$}}}] at ( $ (3)!{1/2}!(5) $ ) (4) {};  %
\draw (5) coordinate[c3,fill=green];  %
\draw (5) coordinate[c2,fill=blue,label=below right:{\footnotesize   \colorbox{white}{$\{2\}$}}];  %
\node[c3,fill=green,label={right:\colorbox{white}{\footnotesize  $\{3,4\}$}}] at ( $ (5)!{1/2}!(1) $ ) (6) {};  %
\end{tikzpicture}
&
\begin{tikzpicture}  [scale=0.2]

\newdimen\ms
\ms=0.05cm

\tikzstyle{every path}=[line width=1pt]

\tikzstyle{c3}=[circle,inner sep={\ms/8},minimum size=3*\ms]
\tikzstyle{c2}=[circle,inner sep={\ms/8},minimum size=1.7*\ms]
\tikzstyle{c1}=[circle,inner sep={\ms/8},minimum size=0.8*\ms]

\newdimen\R
\R=6cm     



\path
  ({45 + 0 * 360 /4}:\R      ) coordinate(1)
  ({45 + 1 * 360 /4}:\R   ) coordinate(3)
  ({45 + 2 * 360 /4}:\R  ) coordinate(5)
  ({45 + 3 * 360 /4}:\R  ) coordinate(7)
;


\draw [color=orange] (1) -- (3);
\draw [color=blue]  (3) -- (5);
\draw [color=red] (5) -- (7);
\draw [color=green] (7) -- (1);

%
%

\draw (1) coordinate[c3,fill=orange];   %
\draw (1) coordinate[c2,fill=green,label=45:{\footnotesize $\{1,3\}$}];  %
%
\node[c3,fill=orange,label={above:{\footnotesize  $\{ 4,5,7\}$}}] at ( $ (1)!{1/2}!(3) $ ) (2) {};    %
\draw (3) coordinate[c3,fill=blue];  %
\draw (3) coordinate[c2,fill=orange,label={180-45}:{\footnotesize {$\{2,6 \}$}}];  %
%
\node[c3,fill=blue,label={left:{\footnotesize {$\{ 3,4,7 \}$}}}] at ( $ (3)!{1/2}!(5) $ ) (4) {};  %
\draw (5) coordinate[c3,fill=red];  %
\draw (5) coordinate[c2,fill=blue,label={180+45}:{\footnotesize  {$\{1,5 \}$}}];  %
%
\node[c3,fill=red,label={below:{\footnotesize $\{ 3,6,7 \}$}}] at ( $ (5)!{1/2}!(7) $ ) (6) {};  %
\draw (7) coordinate[c3,fill=green];  %
\draw (7) coordinate[c2,fill=red,label= {270+45}:{\footnotesize  {$\{2,4\}$}}];  %
\node[c3,fill=green,label={right:{\footnotesize $\{5,6,7\}$}}] at ( $ (7)!{1/2}!(1) $ ) (8) {};  %
\end{tikzpicture}
\\
(a)&(b)
\\
\begin{tikzpicture}  [scale=0.2]

\newdimen\ms
\ms=0.05cm

\tikzstyle{every path}=[line width=1pt]

\tikzstyle{c3}=[circle,inner sep={\ms/8},minimum size=3*\ms]
\tikzstyle{c2}=[circle,inner sep={\ms/8},minimum size=1.7*\ms]
\tikzstyle{c1}=[circle,inner sep={\ms/8},minimum size=0.8*\ms]

\newdimen\R
\R=6cm     



\path
  ({45 + 0 * 360 /4}:\R      ) coordinate(1)
  ({45 + 1 * 360 /4}:\R   ) coordinate(3)
  ({45 + 2 * 360 /4}:\R  ) coordinate(5)
  ({45 + 3 * 360 /4}:\R  ) coordinate(7)
;


\draw [color=orange] (1) -- (3);
\draw [color=blue]  (3) -- (5);
\draw [color=red] (5) -- (7);
\draw [color=green] (7) -- (1);
\draw [color=violet] (7) -- (3);

%
%

\draw (1) coordinate[c3,fill=orange];   %
\draw (1) coordinate[c2,fill=green,label=45:{\footnotesize $\{1,2\}$}];  %
%
\node[c3,fill=orange,label={above:{\footnotesize  $\{3,4,5\}$}}] at ( $ (1)!{1/2}!(3) $ ) (2) {};    %
\draw (3) coordinate[c3,fill=blue];  %
\draw (3) coordinate[c2,fill=orange,label={180-45}:{\footnotesize {$\{6\}$}}];  %
\draw (3) coordinate[c1,fill=violet];  %
\node[c3,fill=blue,label={left:{\footnotesize {$\{1,3,5\}$}}}] at ( $ (3)!{1/2}!(5) $ ) (4) {};  %
\draw (5) coordinate[c3,fill=red];  %
\draw (5) coordinate[c2,fill=blue,label={180+45}:{\footnotesize  {$\{2,4\}$}}];  %
%
\node[c3,fill=red,label={below:{\footnotesize $\{1,3,6\}$}}] at ( $ (5)!{1/2}!(7) $ ) (6) {};  %
\draw (7) coordinate[c3,fill=green];  %
\draw (7) coordinate[c2,fill=red,label= {270+45}:{\footnotesize  {$\{5\}$}}];  %
\draw (7) coordinate[c1,fill=violet];  %
\node[c3,fill=green,label={right:{\footnotesize $\{3,4,6\}$}}] at ( $ (7)!{1/2}!(1) $ ) (8) {};  %
\node[c3,fill=violet,label={[label distance=-2]above right:{\footnotesize $m$}}] at ( $ (7)!{1/2}!(3) $ ) (9) {};  %
\end{tikzpicture}
&
\begin{tikzpicture}  [scale=0.15]

\newdimen\ms
\ms=0.05cm

\tikzstyle{every path}=[line width=1pt]

\tikzstyle{c3}=[circle,inner sep={\ms/8},minimum size=3*\ms]
\tikzstyle{c2}=[circle,inner sep={\ms/8},minimum size=1.7*\ms]
\tikzstyle{c1}=[circle,inner sep={\ms/8},minimum size=0.8*\ms]

\newdimen\R
\R=6cm     



\path
  ({90 + 0 * 360 /5}:\R      ) coordinate(1)
  ({90 + 36 + 0 * 360 /5}:{\R * sqrt((25+10*sqrt(5))/(50+10*sqrt(5)))}      ) coordinate(2)
  ({90 + 1 * 360 /5}:\R   ) coordinate(3)
  ({90 + 36 + 1 * 360 /5}:{\R * sqrt((25+10*sqrt(5))/(50+10*sqrt(5)))}   ) coordinate(4)
  ({90 + 2 * 360 /5}:\R  ) coordinate(5)
  ({90 + 36 + 2 * 360 /5}:{\R * sqrt((25+10*sqrt(5))/(50+10*sqrt(5)))}  ) coordinate(6)
  ({90 + 3 * 360 /5}:\R  ) coordinate(7)
  ({90 + 36 + 3 * 360 /5}:{\R * sqrt((25+10*sqrt(5))/(50+10*sqrt(5)))}  ) coordinate(8)
  ({90 + 4 * 360 /5}:\R     ) coordinate(9)
  ({90 + 36 + 4 * 360 /5}:{\R * sqrt((25+10*sqrt(5))/(50+10*sqrt(5)))}     ) coordinate(10)
;


\draw [color=orange] (1) -- (2) -- (3);
\draw [color=red] (3) -- (4) -- (5);
\draw [color=green] (5) -- (6) -- (7);
\draw [color=blue] (7) -- (8) -- (9);
\draw [color=magenta] (9) -- (10) -- (1);    %
\draw [color=violet] (1) -- (6) ;    %

%
%
\draw (1) coordinate[c3,fill=orange,label=90:{\footnotesize $\{1,2\}$}];   %
\draw (1) coordinate[c2,fill=magenta];  %
\draw (1) coordinate[c1,fill=violet];  %
\draw (2) coordinate[c3,fill=orange,label={above left:\footnotesize $\{3,5,6,8,10\}$}];    %
\draw (3) coordinate[c3,fill=red,label={left:\footnotesize $\{4,7,9\}$}]; %
\draw (3) coordinate[c2,fill=orange];  %
\draw (4) coordinate[c3,fill=red,label={left:\footnotesize $\{2,3,6,8\}$}];  %
\draw (5) coordinate[c3,fill=green,label={below left:\footnotesize $\{1,5,10\}$}];  %
\draw (5) coordinate[c2,fill=red];  %
\draw (6) coordinate[c3,fill=green,label={below:\footnotesize $\{3,4,8,9\}$}];
\draw (6) coordinate[c2,fill=violet];  %
\draw (7) coordinate[c3,fill=blue,label={below right:\footnotesize $\{2,6,7\}$}];  %
\draw (7) coordinate[c2,fill=green];  %
\draw (8) coordinate[c3,fill=blue,label={right:\footnotesize $\{1,3,4,5\}$}];  %
\draw (9) coordinate[c3,fill=magenta,label={right:\footnotesize $\{8,9,10\}$}];
\draw (9) coordinate[c2,fill=blue];  %
\draw (10) coordinate[c3,fill=magenta,label={above right:\footnotesize $\{3,4,5,6,7\}$}];  %
\node[c3,fill=violet,label={[label distance=-2]right:{\footnotesize $m$}}] at ( $ (1)!{1/2}!(6) $ ) (11) {};  %

\end{tikzpicture}
\\
(c)&(d)
\\
\begin{tikzpicture}  [scale=0.15]

\newdimen\ms
\ms=0.05cm

\tikzstyle{every path}=[line width=1pt]

\tikzstyle{c5}=[circle,inner sep={\ms/8},minimum size=5*\ms]
\tikzstyle{c4}=[circle,inner sep={\ms/8},minimum size=4*\ms]
\tikzstyle{c3}=[circle,inner sep={\ms/8},minimum size=3*\ms]
\tikzstyle{c2}=[circle,inner sep={\ms/8},minimum size=1.7*\ms]
\tikzstyle{c1}=[circle,inner sep={\ms/8},minimum size=0.8*\ms]

\newdimen\R
\R=6cm     



\path
  ({90 + 0 * 360 /5}:\R      ) coordinate(1)
  ({90 + 36 + 0 * 360 /5}:{\R * sqrt((25+10*sqrt(5))/(50+10*sqrt(5)))}      ) coordinate(2)
  ({90 + 1 * 360 /5}:\R   ) coordinate(3)
  ({90 + 36 + 1 * 360 /5}:{\R * sqrt((25+10*sqrt(5))/(50+10*sqrt(5)))}   ) coordinate(4)
  ({90 + 2 * 360 /5}:\R  ) coordinate(5)
  ({90 + 36 + 2 * 360 /5}:{\R * sqrt((25+10*sqrt(5))/(50+10*sqrt(5)))}  ) coordinate(6)
  ({90 + 3 * 360 /5}:\R  ) coordinate(7)
  ({90 + 36 + 3 * 360 /5}:{\R * sqrt((25+10*sqrt(5))/(50+10*sqrt(5)))}  ) coordinate(8)
  ({90 + 4 * 360 /5}:\R     ) coordinate(9)
  ({90 + 36 + 4 * 360 /5}:{\R * sqrt((25+10*sqrt(5))/(50+10*sqrt(5)))}     ) coordinate(10)
;


\draw [color=orange] (1) -- (2) -- (3);
\draw [color=red] (3) -- (4) -- (5);
\draw [color=green] (5) -- (6) -- (7);
\draw [color=blue] (7) -- (8) -- (9);
\draw [color=magenta] (9) -- (10) -- (1);    %
\draw [color=violet] (1) -- (6) ;    %
\draw [color=brown] (9) -- (6) ;    %
\draw [color=pink] (3) -- (6) ;    %

%
%
\draw (1) coordinate[c3,fill=orange,label=90:{\footnotesize $\{1,2\}$}];   %
\draw (1) coordinate[c2,fill=magenta];  %
\draw (1) coordinate[c1,fill=violet];  %
\draw (2) coordinate[c3,fill=orange,label={above left:\footnotesize $\{3,4,5,7\}$}];    %
\draw (3) coordinate[c4,fill=red,label={left:\footnotesize $\{6\}$}]; %
\draw (3) coordinate[c3,fill=pink];  %
\draw (3) coordinate[c2,fill=orange];  %
\draw (4) coordinate[c3,fill=red,label={left:\footnotesize $\{2,3,5\}$}];  %
\draw (5) coordinate[c3,fill=green,label={below left:\footnotesize $\{1,4,7\}$}];  %
\draw (5) coordinate[c2,fill=red];  %
\draw (6) coordinate[c5,fill=pink,label={below:\footnotesize $\{3\}$}];
\draw (6) coordinate[c4,fill=brown];
\draw (6) coordinate[c3,fill=green];
\draw (6) coordinate[c2,fill=violet];  %
\draw (7) coordinate[c3,fill=blue,label={below right:\footnotesize $\{2,5,6\}$}];  %
\draw (7) coordinate[c2,fill=green];  %
\draw (8) coordinate[c3,fill=blue,label={right:\footnotesize $\{1,3,4\}$}];  %
\draw (9) coordinate[c4,fill=magenta,label={right:\footnotesize $\{7\}$}];
\draw (9) coordinate[c3,fill=brown];  %
\draw (9) coordinate[c2,fill=blue];  %
\draw (10) coordinate[c3,fill=magenta,label={above right:\footnotesize $\{3,4,5,6\}$}];  %
\node[c3,fill=violet,label={[label distance=-3]left:{\footnotesize $m_3$}}] at ( $ (9)!{1.2/3}!(6) $ ) (11) {};  %
\node[c3,fill=brown,label={[label distance=-4]below right:{\footnotesize $m_2$}}] at ( $ (1)!{1/4}!(6) $ ) (12) {};  %
\node[c3,fill=pink,label={[label distance=-3]right:{\footnotesize $m_1$}}] at ( $ (3)!{1.2/3}!(6) $ ) (13) {};  %

\end{tikzpicture}
&
\begin{tikzpicture}  [scale=0.15]

\newdimen\ms
\ms=0.05cm

\tikzstyle{every path}=[line width=1pt]

\tikzstyle{c4}=[circle,inner sep={\ms/8},minimum size=4*\ms]
\tikzstyle{c3}=[circle,inner sep={\ms/8},minimum size=3*\ms]
\tikzstyle{c2}=[circle,inner sep={\ms/8},minimum size=1.7*\ms]
\tikzstyle{c1}=[circle,inner sep={\ms/8},minimum size=0.8*\ms]

\newdimen\R
\R=6cm     



\path
  ({90 + 0 * 360 /5}:\R      ) coordinate(1)
  ({90 + 36 + 0 * 360 /5}:{\R * sqrt((25+10*sqrt(5))/(50+10*sqrt(5)))}      ) coordinate(2)
  ({90 + 1 * 360 /5}:\R   ) coordinate(3)
  ({90 + 36 + 1 * 360 /5}:{\R * sqrt((25+10*sqrt(5))/(50+10*sqrt(5)))}   ) coordinate(4)
  ({90 + 2 * 360 /5}:\R  ) coordinate(5)
  ({90 + 36 + 2 * 360 /5}:{\R * sqrt((25+10*sqrt(5))/(50+10*sqrt(5)))}  ) coordinate(6)
  ({90 + 3 * 360 /5}:\R  ) coordinate(7)
  ({90 + 36 + 3 * 360 /5}:{\R * sqrt((25+10*sqrt(5))/(50+10*sqrt(5)))}  ) coordinate(8)
  ({90 + 4 * 360 /5}:\R     ) coordinate(9)
  ({90 + 36 + 4 * 360 /5}:{\R * sqrt((25+10*sqrt(5))/(50+10*sqrt(5)))}     ) coordinate(10)
;


\draw [color=orange] (1) -- (2) -- (3);
\draw [color=red] (3) -- (4) -- (5);
\draw [color=green] (5) -- (6) -- (7);
\draw [color=blue] (7) -- (8) -- (9);
\draw [color=magenta] (9) -- (10) -- (1);    %
\draw [color=violet] (1) -- (6) ;    %
\draw [color=brown] (5) -- (9) ;    %

%
%
\draw (1) coordinate[c3,fill=orange,label=90:{\footnotesize$\{1\}$}];   %
\draw (1) coordinate[c2,fill=magenta];  %
\draw (1) coordinate[c1,fill=violet];  %
\draw (2) coordinate[c3,fill=orange,label={above left:\footnotesize $\{2,4\}$}];    %
\draw (3) coordinate[c3,fill=red,label={left:\footnotesize $\{3,5\}$}]; %
\draw (3) coordinate[c2,fill=orange];  %
\draw (4) coordinate[c3,fill=red,label={left:\footnotesize $\{2,4\}$}];  %
\draw (5) coordinate[c4,fill=green,label={below left:\footnotesize $\{1\}$}];  %
\draw (5) coordinate[c3,fill=brown];  %
\draw (5) coordinate[c2,fill=red];  %
\draw (6) coordinate[c3,fill=green,label={below:\footnotesize $\{4,5\}$}];
\draw (6) coordinate[c2,fill=violet];  %
\draw (7) coordinate[c3,fill=blue,label={below right:\footnotesize $\{2,3\}$}];  %
\draw (7) coordinate[c2,fill=green];  %
\draw (8) coordinate[c3,fill=blue,label={right:\footnotesize $\{1\}$}];  %
\draw (9) coordinate[c4,fill=magenta,label={right:\footnotesize $\{4,5\}$}];
\draw (3) coordinate[c3,fill=brown];  %
\draw (9) coordinate[c2,fill=blue];  %
\draw (10) coordinate[c3,fill=magenta,label={above right:\footnotesize $\{2,3\}$}];  %
\node[c3,fill=violet,label={[label distance=-4]above left:{\footnotesize $m$}}] at ( $ (5)!{7.7/20}!(9) $ ) (11) {};  %
\draw (11) coordinate[c2,fill=brown];  %

\end{tikzpicture}
\\(e)&(f)
\end{tabular}
\end{center}
\caption{
Greechie orthogonality diagram of partition logic realizations of
(a) Wright's triangle logic with a partition logic realization~\cite{wright,dvur-pul-svo});
(b) a square logic from the cyclic pasting of four three-atomic edges/cliques/contexts $K_3$;
(c) a combo of triangle logics pasted along a common edge with a partition logic realization~\cite[Example~8.3, pp.~ 421,422]{dvur-pul-svo}, with $m=\{1,2,3,4\}$;
(d) pentagon logic with inner edge, with a partition logic realization, and $m=\{5,6,7,10\}$;
(e) pentagon logic with three inner edges,  with a partition logic realization, and
$m_1=\{1,2,4,5,7\}$,
$m_2=\{4,5,6,7\}$,
$m_3=\{1,2,4,5,6\}$;
(e) pentagon logic with two inner edges and $m=\{2,3\}$, without a partition logic realization
since the vertex representation generated from the 5 dispersion free weights is highly degenerate and nonseparating.
}
\label{2018-q-f2}
\end{figure}
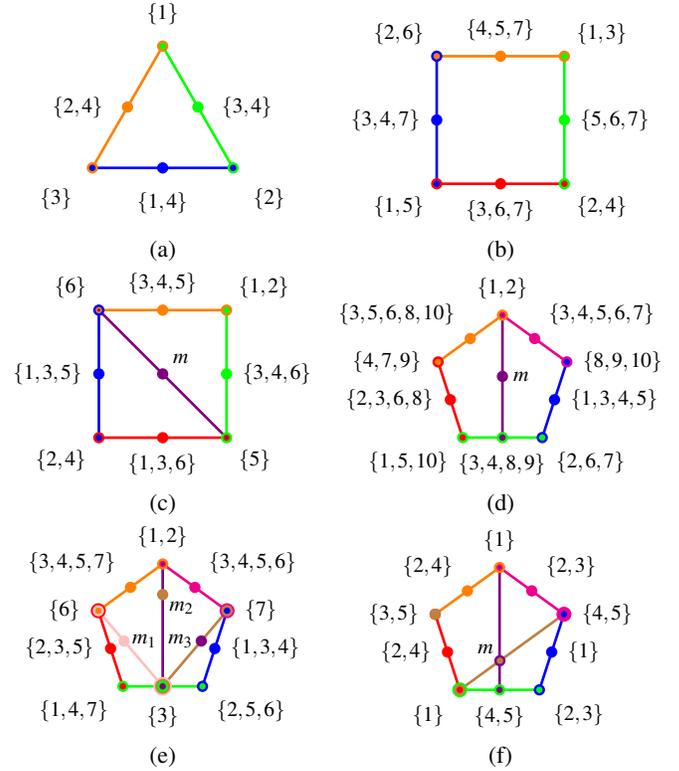

Let us, for a moment, consider partition logics not restraint by low-dimensional faithful orthogonal representability,
but with a separable set of two-valued states [with the exception of the logic depicted in Figure~\ref{2018-q-f2}(f)].
These have no quantum realization. Yet, due to the automaton logic, they are intrinsically realized in Turing universal environments.

An example of such a structure is Wright's triangle logic~\cite[Figure~2, p.~900]{wright} depicted in Figure~\ref{2018-q-f2}(a)
(see also~\cite[Fig.~6, Example~8.2, pp.~414,420,421]{dvur-pul-svo}).
As mentioned earlier, together with 4 dispersion free weights it allows another weight $\frac{1}{2}$ on its intertwining vertices.
Figures~\ref{2018-q-f2}(b{\&}c) depict a square logic, the latter one being formed by a pasting of two triangle logics along a common leg.
Figures~\ref{2018-q-f2}(d{\&}e) depict pentagon logic with one and three inner cliques/contexts;
the latter one realizing a true-implies-four-times-true configuration for $\{3\}$.
Figure~\ref{2018-q-f2}(f) has no partition logic representation, as its 5 dispersion free weights cannot separate three atoms realized by $\{1\}$ and $\{2,3\}$,
as well as two atoms realized by $\{2,4\}$ and $\{4,5\}$, respectively.

\section{Identical graphs realizable by different physical resources require different probability types}

It is important to emphasize that both scenarios -- the classical generalized urn scenario as well as the quantized one --
from a graph-theoretical point of view, operate with identical (exclusivity) graphs
(e.g., Figures~1 in both References~\cite{fritz-2013,Cabello-2014-gtatqc}).
The difference is the representation of these graphs: the quantum case has a faithful orthogonal representation in some finite-dimensional Hilbert space,
whereas the classical case in terms of a generalized urn model has a
set-theoretic representation in terms of partitions of some finite set.

Generalized urn models and automaton logics are models of partition logics which
are capable of complementarity yet fail to render (quantum) value indefiniteness.
They are important for an understanding of the ``twilight zone'' spanned by nonclassicality
(nondistributivity, nonboolean logics) and yet full value definiteness -- one may call this a ``purgatory'' -- floating in-between classical Boolean and quantum realms.

It should be stressed that the algebraic structure of empirical logics, or graphs, do in general {\em not} determine the
types of probability measures on them.
For instance, a generalized urn loaded with balls rendering the pentagon structure, as envisioned by Wright,
has probabilities different from the scheme of Gr{\"o}tschel, Lov{\'a}sz and  Schrijver, which is based on orthogonal representations of the pentagon.
Likewise, a geometric resource such as a ``vector contained in a box'' and ``measured along projections onto an orthonormal basis''
will not conform to probabilities induced by the convex hull of the dispersion-free weights -- even if these weights
are separating.
Therefore the particular physical resource -- what is actually inside the black box -- determines which type of probability theory is applicable.

Furthermore, partition logics which are not just a single Boolean algebra represent empirical configurations featuring complementarity.
And yet they all have separating~\cite[Theorem~0]{kochen1} sets of two-valued states and thus are not ``contextual'' in the Specker sense~\cite{specker-60}.



\begin{acknowledgments}
This work was greatly inspired by Ad\'an Cabello's insistence on graph theoretical importance for quantum mechanics, and for his challenge to come up with a classical
Einstein-Podolsky-Rosen-type scenario.
\end{acknowledgments}


%

\end{document}